\documentclass[aps,prb,reprint,superscriptaddress,amsmath,amssymb,floatfix]{revtex4-2}
\usepackage{graphicx}

\usepackage{siunitx}
\usepackage{hyperref}
\hypersetup{  colorlinks=true,     urlcolor=blue,     citecolor=blue,     linkcolor=blue,}
\def\cm{cm$^{-1}$}
\def\Cl{$\kappa$-Cl}
\def\Br{$\kappa$-Br}
\def\I{$\kappa$-I}
\def\X{$\kappa$-(BEDT-TTF)$_2$Cu[N(CN)$_2$]$X$}

\begin{document}

\title{Charge localization in strongly correlated $\kappa$-(BEDT-TTF)$_2$Cu[N(CN)$_2$]I\\
due to inherent disorder}

\author{O.~Iakutkina}
\email{olga.iakutkina@pi1.uni-stuttgart.de}
\affiliation{1.~Physikalisches Institut, Universit{\"a}t Stuttgart, 70569 Stuttgart, Germany}

\author{L.N.~Majer}
\affiliation{1.~Physikalisches Institut, Universit{\"a}t Stuttgart, 70569 Stuttgart, Germany}

\author{T.~Biesner}
\affiliation{1.~Physikalisches Institut, Universit{\"a}t Stuttgart, 70569 Stuttgart, Germany}

\author{E. Uykur}
\affiliation{1.~Physikalisches Institut, Universit{\"a}t Stuttgart, 70569 Stuttgart, Germany}

\author{J.A.~Schlueter}
\affiliation{Material Science Division, Argonne National Laboratory, Argonne, IL 60439-4831 and \\
National Science Foundation, Alexandria, VA 2223, U.S.A.}

\author{M.~Dressel}
\affiliation{1.~Physikalisches Institut, Universit{\"a}t Stuttgart, 70569 Stuttgart, Germany}

\date{\today}

\begin{abstract}
In order to understand the physical properties of the series of organic conductors $\kappa$-(BEDT-TTF)$_2$Cu[N(CN)$_2$]$X$ with $X$ = Cl, Br, and I, not only electronic correlations but also the effect of disorder has to be taken into account. While for Cl- and Br-containing salts the influence of both parameters were investigated and a universal phase diagram was proposed, the position of $\kappa$-(BEDT-TTF)$_2$Cu[N(CN)$_2$]I is still not settled. Here we have conducted transport, infrared, and dielectric measurements on single crystals of the title compound to clarify its electronic state at low temperatures. The correlation strength was determined as $U/W \approx 2.2$; thus this salt is placed deeper in an insulating state compare to the two sister compounds. We found that inherent disorder leads to a Coulomb localized insulating state similar to the moderately x-ray-irradiated $\kappa$-(BEDT-TTF)$_2$Cu[N(CN)$_2$]Cl.
\end{abstract}

\maketitle

\section{\label{sec:introduction}Introduction}

The Mott metal-insulator transition is one of the most intriguing phenomena in condensed matter physics \cite{Mott} that is not completely understood yet. Here the insulating state arises from strong on-site Coulomb repulsion $U$ with respect to the bandwidth $W$. Organic charge-transfer salts are perfectly suited systems for investigating  such metal-insulator transition, as they can be easily tuned through the transition both by chemical substitution, or by applying relatively weak hydrostatic pressure. Among them the $\kappa$-phase BEDT-TTF compounds have attracted great attention as a bandwidth-controlled Mott system with a variety of exotic ground states \cite{Miyagawa2004,PhysRevB.76.165113,*PhysRevB.79.195106,Yasin2011,Dressel20}.

The family of isostructural salts $\kappa$-(BEDT-TTF)$_2$\-Cu[N(CN)$_2$]$X$ (where BEDT-TTF denotes bis\-(ethylene\-dithio)tetrathiafulvalene and $X$ = Cl, Br, I; here abbreviated as $\kappa$-Cl, $\kappa$-Br, and $\kappa$-I, respectively) was discovered in the early 1990s by J.M. Williams and collaborators \cite{WANG19911983, GEISER1991475}. By increasing the halogen atom size, the system is expected to go from an insulating to a metallic state. Indeed, the first two salts follow this dependence: \Cl\ is a dimer Mott insulator that becomes superconducting by applying a tiny hydrostatic pressure of only 0.3~kbar ($T_c = 12.8$~K) \cite{WANG19911983, Williams1990}. When Cl is replaced by the bigger Br atom, \Br\ shows metallic and even superconducting behavior ($T_c \approx 11$~K) already at ambient pressure \cite{Kini1990}. However, when we go to the even bigger halogen atom - I, the trend does not continue: \I\ is a para\-magnetic insulator at ambient pressure and becomes superconducting only under much higher pressures compare to \Cl\ ($T_c = 8$~K at $p \approx 1.2$~kbar depending on the quality of crystals) \cite{PhysRevB.62.15561,Kushch2001,Kushch2003}. Previous studies inferred that in this case the insulating state arises from superlattice formation in the anion layer \cite{PhysRevB.62.15561}. In addition, it was shown that in \I\ inherent disorder plays an important role and should be considered when describing the low-temperature state \cite{PhysRevB.100.195115,Tsuchiya2020,PhysRevB.102.214430,GEISER1991475}. For the other two members of the family, \Cl\ and \Br, it was recently recognized that disorder --~introduced by x-ray irradiation~-- severely alters the physical properties \cite{cryst2020374,PhysRevLett.101.206403, PhysRevLett.104.217003, Sasaki2012}. Therefore, to describe completely all three  \X\ salts, not only electronic correlations but also disorder has to be considered.

The comprehensive investigation of the physical properties of \I\ by transport measurement, infrared and dielectric spectroscopies, presented here, enables us now to put \I\ in line with the other two members of the \X\ family.

\section{\label{sec:metods}Materials and methods}

\begin{figure}[h]
    \centering\includegraphics[width=1\columnwidth]{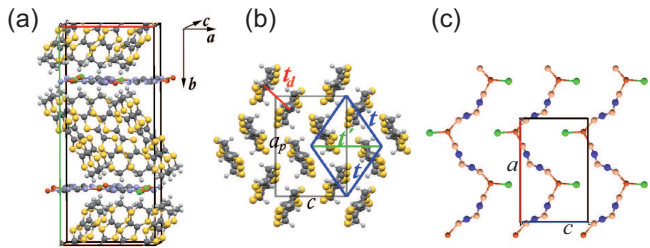}
	\caption{Crystal structure of $\kappa$-(BEDT-TTF)$_2$Cu[N(CN)$_2$]I. The lines mark the unit cell. Carbon, sulfur and hydrogen atoms of the BEDT-TTF molecule are colored in dark gray, yellow and light gray;
for the anion chains, iodine, cooper, carbon and nitrogen are colored in green, red, blue and orange, respectively.
(a) The layers of BEDT-TTF molecules are separated by planes of anions along the $b$-direction. The alternating tilting direction of the BEDT-TTF dimers in adjacent layers leads to a doubling of the unit cell.
In panels (b) and (c) one cation and anion layer is shown, respectively, where $a_p$ denotes the projection of the $a$ axis on the direction perpendicular $b$ and $c$ axes.  The interdimer transfer integrals are denoted by $t$ and $t^{\prime}$, and the intradimer transfer integral by $t_\mathrm{d}$. }
	\label{fig:structure}
\end{figure}

Single crystals of $\kappa$-(BEDT-TTF)$_2$Cu[N(CN)$_2$]I were grown at the Argonne National Laboratory by standard electrochemical oxidation method according to the procedure described in Ref.~\onlinecite{Kushch2003}. The
crystals have a quasi-two-dimensional structure composed by alternating BEDT-TTF donor layers,
separated along the $b$-direction by  insulating Cu[N(CN)$_2$]I$^-$ sheets, as depicted in Fig.~\ref{fig:structure}.
The polymeric anionic zig-zag chains extend along the $a$-axis.
The BEDT-TTF molecules compose dimers that are tilted with respect to the $b$-axis, forming alternating layers
in a herring bone fashion. Their inclination in $a$-direction results in short contacts between the ethylene end-groups and the anionic chains.

All $\kappa$-(BEDT-TTF)$_2$Cu[N(CN)$_2$]$X$ compounds crystalize in $Pnma$ space group with $Z=4$, i.e. four dimers per unit cell. While at room temperature the $c$-parameter increases with size of the halogen atom, the unit cell shrinks along the $a$-axis; hence
the area $a \times c$ needed to pack four donor molecules remains nearly constant for all three compounds \cite{GEISER1991475}. What is the most important here is the aspect ratio $c/a$, which reflects the effect of chemical pressure on the \X\ salts \cite{mori1999}. Even though the values of $c/a$ are very similar in all three salts for $T=295$~K, the difference becomes significant at low temperatures, as can be seen from Table \ref{table:1}. T. Mori {\it et al.} showed \cite{mori1999} that for $c/a$ in the range from 0.640 to 0.675,
the correlation strength is reduced upon increasing the axes ratio, in accordance with experimental results for \Cl\ and \Br-salts.
When the aspect ratio $c/a$ increases further, the system becomes highly correlated again because the overlap integrals change significantly.
This could be exactly the case for $\kappa$-(BEDT-TTF)$_2$Cu[N(CN)$_2$]I.

\begin{table}[h]
        \caption{Room-temperature unit-cell data for $\kappa$-(BEDT-TTF)$_2$Cu[N(CN)$_2$]$X$ ($X$ = Cl, Br, I) listed
        together with the axes ratio $c/a$ for room temperature and for the lowest $T$ accessible (taken from Ref.~\onlinecite{GEISER1991475})}
\label{table:1}
    \begin{center}
    \begin{tabular}{r| c c c}
      $X =$~~~    & Cl &  Br &  I \\
    \hline
        $a$ (\si{\angstrom})~~~ & ~~~12.977~~~ & ~~~12.942~~~ & ~~~12.928    \\
        $b$ (\si{\angstrom})~~~ & 29.979 & ~~~30.016~~~  & ~~~30.356 \\
        $c$ (\si{\angstrom})~~~ & ~~8.480 & ~~~~8.539~~~ & ~~~~8.683  \\
        $c/a$ ($T=295$~K)~~~ & ~~0.654 & ~~~~0.660~~~ & ~~~~0.672  \\
        $c/a$ ($T=127$~K)~~~ & ~~0.652 & ~~~~0.659~~~ & ~~~~0.685  \\

    \end{tabular}
    \end{center}
\end{table}

The in-plane resistivity along the $c$-axis was measured by dc two-point method as a function of temperature. Optical spectroscopy was performed in a broad frequency range at different temperatures utilizing two Fourier transform infrared spectrometers. For covering the high-frequency range (above 600~\cm) a Bruker Vertex 80v spectrometer with an attached Hyperion IR microscope was used, while for the low-frequency range (70-700~\cm) reflectance measurements were performed with a Bruker IFS113v spectrometer applying an {\it in-situ} freshly evaporated gold overcoating technique. From the reflectivity spectra, the optical conductivity was calculated employing  Kramers-Kronig analysis with constant extrapolation for low-frequency range, and standard $\omega^{-4}$ decay for high frequencies.
The temperature dependence of the complex dielectric function $\hat{\epsilon}= \epsilon^{\prime} + {\rm i} \epsilon^{\prime\prime}$  was measured with the help of an Agilent A4294 impedance analyzer in the frequency range from $100~{\rm Hz}$ to $10~{\rm MHz}$ along the $c$-axis. Using a home-made sample holder enables us to conduct reliable measurements from 5-10~kHz to  5-10~MHz, where the limits depend on the sample resistivity.
\begin{figure}[t]
\centering
\includegraphics[width=1\columnwidth,clip]{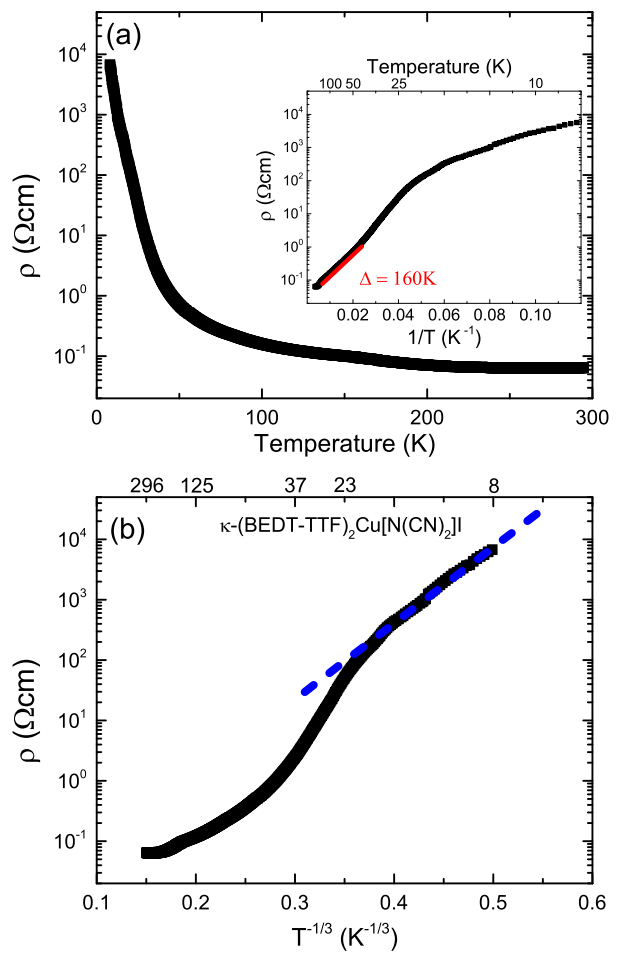}
\caption{Temperature dependence of the dc resistivity of $\kappa$-(BEDT-TTF)$_2$Cu[N(CN)$_2$]I measured in-plane along the $c$-direction. (a)~The log-lin representation visualizes the pronounced insulating behavior at low tempearturs. From the Arrhenius plot in the inset a thermally activated transport can be identified between 50 and 100~K, where the resistivity follows the red line corresponding to a gap $\Delta = 160$~K. (b)~When plotting the resistivity as a function of $T^{-1/3}$, $\rho(T)$ can be fitted by the dashed blue line representing
two-dimensional variable-range-hopping.}
\label{fig:transport}
\end{figure}

\section{\label{sec:results}Results}

\subsection {\label{sec:transport}Transport properties}

Fig.~\ref{fig:transport}(a) displays the dc resistivity $\rho(T)$ of $\kappa$-(BEDT-TTF)$_2$Cu[N(CN)$_2$]I recorded along the $c$-axis
as a function of the temperature. The crystal exhibits a metallic behavior at high $T$ and becomes insulating below approximately 100~K. A transformation to an insulating state happens in the intermediate range between 230 and 100~K, most likely due to a superstructure formation, which doubles the unit cell in the $c$-direction \cite{PhysRevB.62.15561}. Interestingly, a similar formation of a $c^*/2$ superlattice was observed in \Br\ below 200~K while it has not been reported in \Cl\ \cite{WATANABE19991909,NOGAMI1994113}. Tanatar {\it et al.} suggested that the real gap in $\kappa$-(BEDT-TTF)$_2$Cu[N(CN)$_2$]I starts opening only below 100~K due to  short-range ordering with a wave
vector close to $c^*/3$, where the resistivity starts to follow thermally activated behavior \cite{PhysRevB.62.15561}.

The inset of Fig.~\ref{fig:transport} visualizes the resistivity in an Arrhenius plot illustrating the thermally activated behavior of $\rho(T)$ between 100 and 50~K; it starts to deviate from the straight line at low temperatures.
Despite the limited range, we can extract an energy gap $\Delta~\approx~160$~K that is in accord with previous studies \cite{WANG19911983,PhysRevB.62.15561,PhysRevB.102.214430}. Below $T=25$~K the resistivity is significantly reduced compared to a thermally activated behavior. For systems with inherent disorder, electronic transport can take place via hopping between neighboring sites; in this case the temperature dependence of the resistivity follows the variable-range-hopping (VRH) model, described by
\begin{equation}
\rho(T)\propto exp \bigg\{ \frac{T_0}{T} \bigg\}^{1/(1+d)} \quad ,
\label{eq:DChopping}
\end{equation}
where $d=2$ for two-dimensional systems. In Fig.~\ref{fig:transport}(b) the dc resistivity is plotted logarithmically versus $T^{-1/3}$, and indeed it is clearly seen that the plot is linear at low temperatures ($T < 25$~K). This supports the idea of disorder important in \I\ at low temperatures; it is in accord with previous findings \cite{PhysRevB.62.15561, PhysRevB.100.195115}. However, the temperature where the inhomogeneous electronic state appears is slightly lower than was reported previously (40~K). This can be explained by the different quality of crystals.

\subsection{\label{sec:optics}Optical spectroscopy}

\begin{figure}[t]
\centering
\includegraphics[width=1\columnwidth,clip]{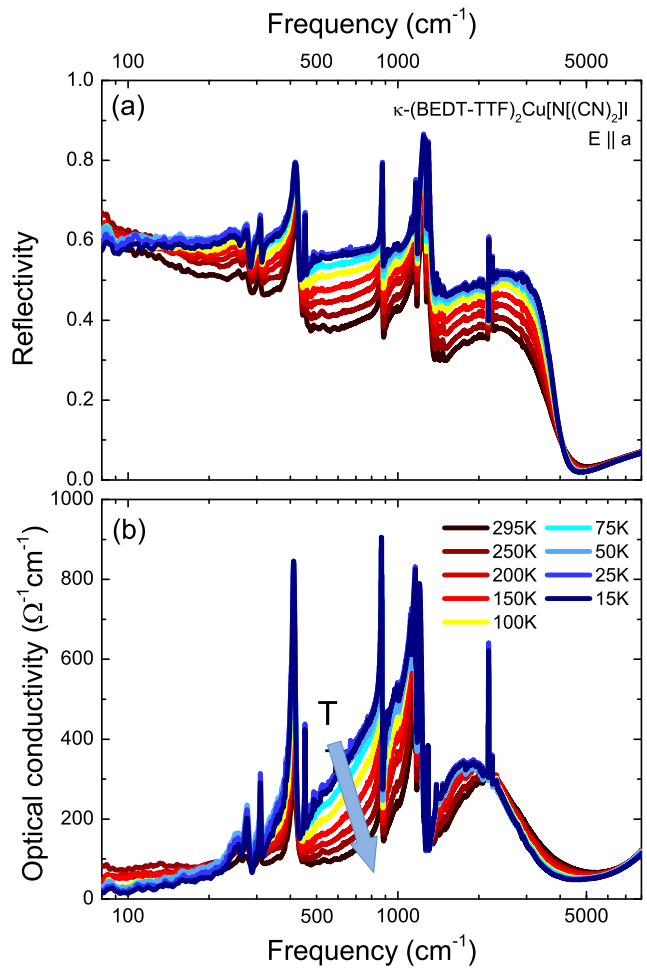}
\caption{(a) Optical reflectivity and (b) conductivity spectra of $\kappa$-(BEDT-TTF)$_2$Cu[N(CN)$_2$]I for $E \parallel  a$ polarization (the most conducting axis) recorded between room temperature and $T=15$~K.}
\label{fig:A}
\end{figure}

\begin{figure}[t]
\centering
\includegraphics[width=1\columnwidth,clip]{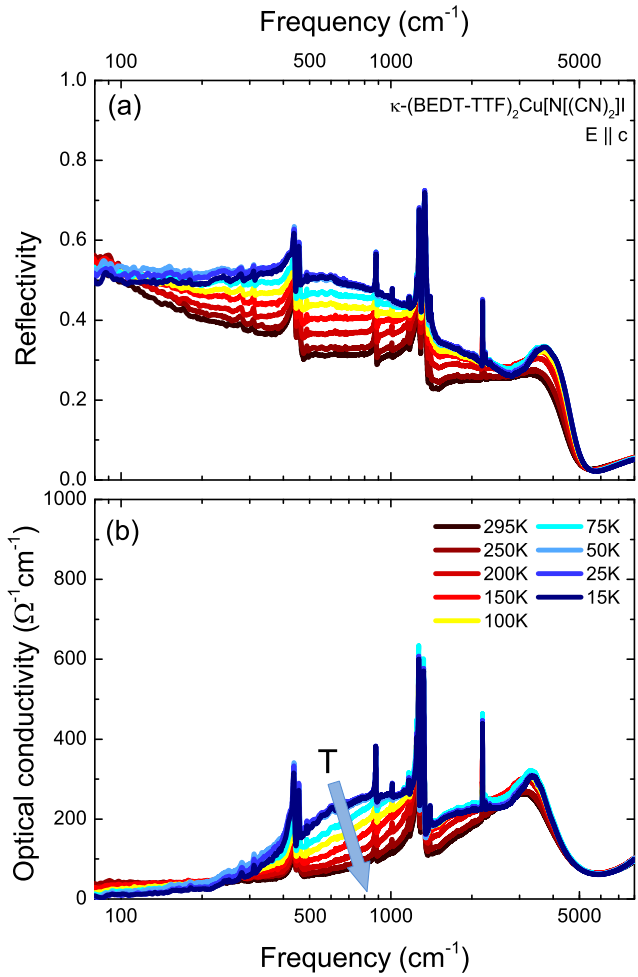}
\caption{(a) In-plane reflectivity and (b) optical conductivity of $\kappa$-(BEDT-TTF)$_2$Cu[N(CN)$_2$]I measured with  $E \parallel c$-axis for temperatures between $T=295$ and 15~K.}
\label{fig:C}
\end{figure}

In Figs.~\ref{fig:A} and \ref{fig:C} the reflectivity and the resulting optical conductivity spectra of $\kappa$-(BEDT-TTF)$_2$Cu[N(CN)$_2$]I are plotted for different temperatures; the light is polarized along the two in-plane directions, i.e.\ $E\parallel a$ and $E \parallel c$. For $E \parallel c$ the infrared spectra are dominated by the large absorption peaks centered around $2200$~\cm\ and $3200$~\cm.
In the perpendicular direction, $E \parallel a$, both transitions coincide in energy, resulting in a single feature.
These bands correspond to intraband transitions between the lower and upper Mott-Hubbard bands and interband between the dimer bands,
respectively and are typical for the $\kappa$-type phase of the  BEDT-TTF salts \cite{PhysRevB.76.165113,*PhysRevB.79.195106,PhysRevB.89.205106}.

In addition to the electronic features, there are strong vibrational signatures at 450, 850 and $1200$~\cm. These molecular vibrations are assigned to $\nu_{10}(a_g)$, $\nu_{49}(b{_{2u}})$ and $\nu_{3}(a_g)$ modes of the BEDT-TTF molecules, respectively \cite{Dressel2004,Girlando2011,Eldridge1996}. The two totally symmetric stretching modes can be observed by infrared spectroscopy 
due to electron-molecular vibration (emv) coupling. In addition, several peaks with smaller intensities are present in conductivity spectra,
which are assigned in the Appendix~\ref{Appendix:Vibrations}; we also refer to comprehensive vibrational studies by Eldridge and others \cite{Eldridge1996,ELDRIDGE1991583,ELDRIDGE1995947}.

At ambient condition, \I\ is a correlated metal, with relatively low reflectivity that continuously drops with frequency; hence the corresponding optical conductivity is rather small. Upon cooling, the crystal becomes insulating with a frequency-independent reflectivity in the far-infrared range; the Mott  gap opens. Reducing the temperature further, a pronounced in-gap absorption in the far-infrared region 
occurs. The same was observed in the irradiated \Cl\ salt, where the introduction of disorder in the anion layer by x-ray radiation leads to a filling of the Mott gap: the Mott insulating state  is transformed to a localization insulating state \cite{cryst2020374,PhysRevLett.101.206403}. The in-gap absorption is more prominent along the $c$-direction where electronic properties are stronger affected by the superstructure formation in the anion layers \cite{PhysRevB.62.15561}, therefore in the discussion we will focus 
on infrared spectra for the polarization $E \parallel c$.

In addition, to make a full comparison of all three \X\ ($X$= Cl, Br, I) salts, the in-plane reflectivity of \Br\ was measured for $E \parallel (a,c)$ at $T = 20$~K, and the optical conductivity was subsequently calculated via Kramers-Kronig transformation. The results will be discussed in Sec.~\ref{sec:Discussion}.

\subsection{\label{sec:dielectrics}Dielectric spectroscopy}

\begin{figure}[t]
\centering
\includegraphics[width=1\columnwidth,clip]{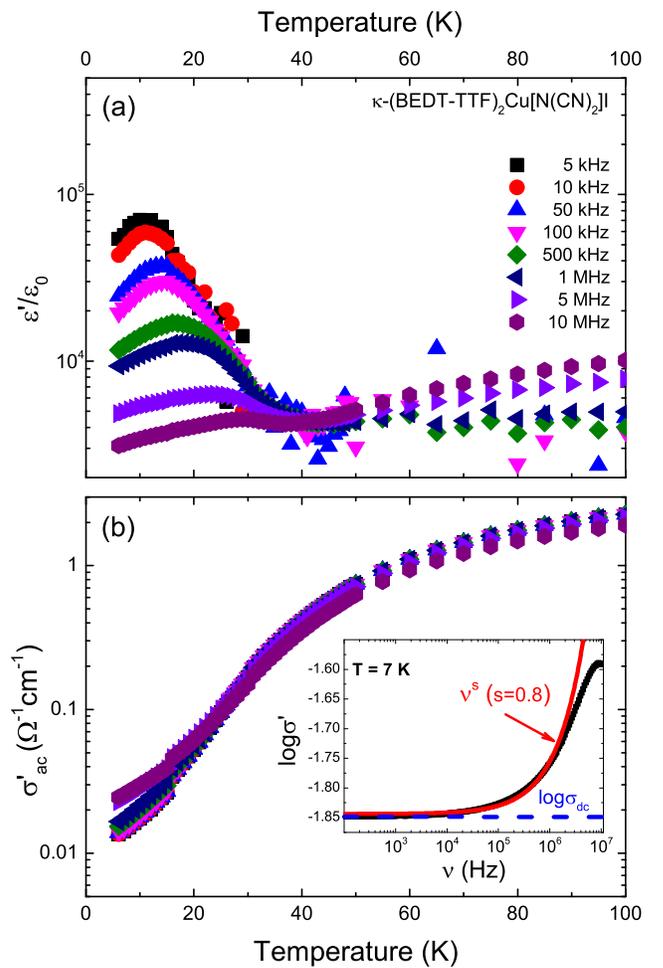}
\caption{Temperature dependence of (a) the real part of dielectric constant $\epsilon^{\prime}(\omega)/\epsilon_0$ and (b) the real part of the ac conductivity $\sigma^{\prime}(\omega)$ of $\kappa$-(BEDT-TTF)$_2$Cu[N(CN)$_2$]I for  $E \parallel c$-axis measured at different frequencies $\nu$ as indicated.
In the insert, the frequency dependence of the low-temperature conductivity $\sigma^{\prime}(\omega)$ is shown, where the red line corresponds to a variable-range-hopping fit with the critical exponent $s = 0.8$ and the offset is equal to the dc contribution $\sigma_{dc}$.}
\label{fig:dielectric}
\end{figure}

The temperature dependence of the dielectric constant is plotted in Fig.~\ref{fig:dielectric}(a)
for various frequencies as indicated. As data for 5 and 10 kHz are quite noisy above 30~K and not displayed. A relaxor-like anomaly is observed below approximately 40~K with the maximum around 10~K in the static limit;
the peak in $\epsilon^{\prime}(T)$ shifts to higher temperatures as the frequency increases.
The ac conductivity $\sigma^{\prime}=\omega\epsilon^{\prime\prime}/4\pi$ also exhibits a frequency dependence that becomes significant below 20 to 25~K as displayed in Fig.~\ref{fig:dielectric}(b). This relaxational behavior is widely observed in disordered systems such as glass-forming liquids, spin-/cluster glasses, and relaxor ferroelectrics.

In electronic conductors, sufficient disorder can prevent coherent metallic transport and localize charge carriers; nevertheless
charge transport can takes place via hopping between discrete sites. Mott's VRH model gives a theoretical treatment of hopping transport; in Sec. \ref{sec:transport} it was already successfully applied  to describe the temperature-dependent dc conductivity. For the frequency-dependence of the real part of the complex $\hat{\sigma}(\omega)$, hopping conduction results in a power law with an exponent $s < 1$, according to \cite{Jonscher1983,Jonscher1999}:
\begin{equation}
\sigma'=\sigma_{dc}+\sigma_0\nu^s \quad ,
\label{eq:VRD}
\end{equation}
where $\sigma_{dc}$ denotes the dc conductivity, $\sigma_0$ is a prefactor, and $\nu=\omega/2\pi$ is the applied frequency. From the VRH model  $s \approx 0.8$ is expected \cite{Elliott1987,Mott2012,RevModPhys.72.873}.
Indeed, as it can be seen from the inset of Fig.~\ref{fig:dielectric}(b),  $\sigma'(\nu)$ can be well described by Eq.~(\ref{eq:VRD}) with  $s \approx 0.8$ for frequencies below 1~MHz; the deviation at high frequencies may be caused by a weak dependence of $s$ on frequency.
It is interesting to compare these finding with the quantum spin liquid candidates  $\kappa$-(BEDT-TTF)$_2$Cu$_2$(CN)$_3$ and $\kappa$-(BEDT-TTF)$_2$Ag$_2$(CN)$_3$, where exponents $s \approx 0.8 - 1.2$ were reported at low temperatures  \cite{PhysRevLett.121.056402,Dressel18}.

\section{\label{sec:Discussion}Discussion}

\subsection{\label{sec:electronic correlations}Electronic correlations}
For the family of \X\ ($X$ = Cl, Br, I) both, electronic correlations and disorder
are extremely important for understanding the electronic properties;
the former one can lead to the Mott metal-insulator transition, the latter one may result  in Anderson localization. 
Sasaki {\it et al.} previously showed \cite{PhysRevLett.101.206403,cryst2020374}
that increasing disorder by successive x-ray irradiation does not affect the fundamental electronic parameters such as on-site Coulomb repulsion $U$ and bandwidth $W$. Since the correlation strength $U/W$ remains basically unaffected, for all three compounds it can be determined from pristine crystals.

\begin{figure}[t]
\centering
\includegraphics[width=1\columnwidth,clip]{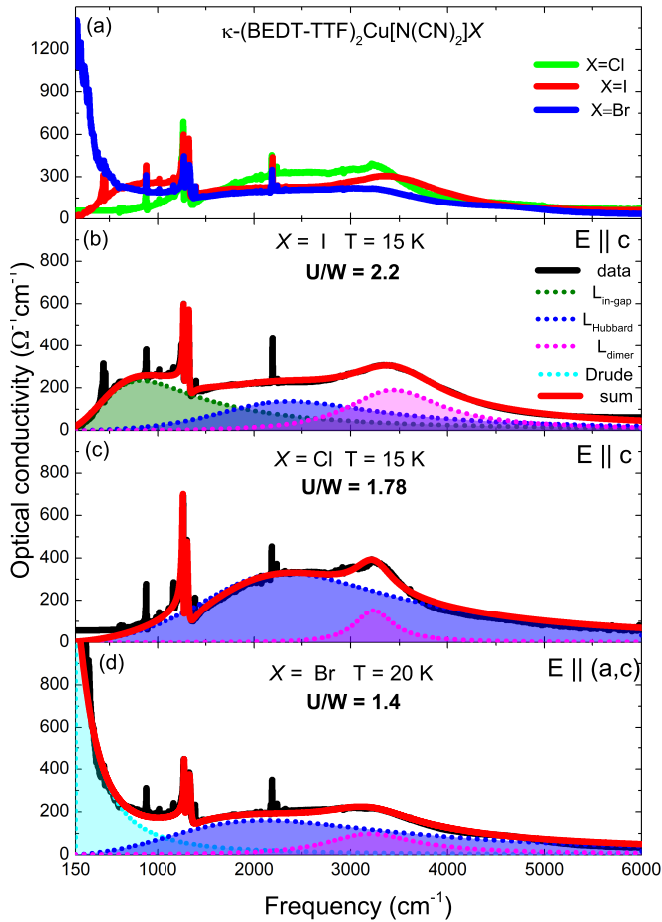}
\caption{(a) Frequency dependence of the optical conductivity of three \X\ salts (X=I, Cl, Br) at the lowest accessible temperatures shown on one graph for comparison. (b) Spectrum of $\kappa$-(BEDT-TTF)$_2$Cu[N(CN)$_2$]I at $T=15$~K and with $E\parallel c$. (c) Spectrum of $\kappa$-(BEDT-TTF)$_2$Cu[N(CN)$_2$]Cl at $T=15$~K and with $E \parallel c$, the data are taken from Ref. \onlinecite{PhysRevB.86.245103}. (d) Spectrum of $\kappa$-(BEDT-TTF)$_2$Cu[N(CN)$_2$]Br at $T=20$~K and with $E \parallel (a,c)$-plane. All spectra were fitted according to the Drude-Lorentz model with indication of different contributions. The red lines correspond to the sum.}
\label{fig:correlations}
\end{figure}

A fit of the optical conductivity by the Drude-Lorentz model has previously \cite{Dressel2004,Dressel2002} been established as a reliable method to determine $U/W$. Following this approach we can  separate the contributions of conduction electrons, interband transitions, and vibrational features. For comparison, the low-temperature optical conductivity of all three salts are displayed in Fig.~\ref{fig:correlations}(a). All data were recorded within the highly conducting $(a,c)$ plane, for \I\ and \Cl\ the polarization $E\parallel c$ is specified.
While the Br-compound is a metal with a prominent Drude-like contribution, \Cl\ is insulating with the Mott gap below approximately 1000~\cm.
\I\ is an insulator, too, but some pronounced in-gap absorption is present for frequencies below 1000~\cm.
For the title compound, inherent disorder leads to a localized insulating state.
In general, Mott insulators exhibit a clear cut gap with no density of states at the Fermi level;
however, here a finite density of states extends close to $E_F$ \cite{cryst2020374}. We should recall that the compound is a clear-cut insulator as far as the dc conductivity is concerned, displayed in Fig.~\ref{fig:transport}.

On the first glance, a similar observation was reported for $\kappa$-(BEDT-TTF)$_2$Cu$_2$(CN)$_3$
\cite{PhysRevB.74.201101,PhysRevB.86.155150,Pustogow2018}, where a rising in-gap absorption upon cooling could finally be explained by entering the phase-coexistence regime linked to the first-order nature of the Mott transition. The enhanced conductivity corresponds to an enormous peak in the dielectric permittivity due to the percolative nature of the metal-insulator transition \cite{Pustogow2021a,PhysRevB.103.125111}. A closer look, however, reveals distinct differences in the spectral and temperature behavior observed in \I. Hence, we conclude fundamentally different reasons for the appearance of excess absorption in these Mott insulators.

An enhancement of the conductivity was also reported for \Cl\ after x-ray irradiation \cite{PhysRevLett.101.206403}.
To clarify this point, we compare the low-temperature optical conductivity of the three sibling compounds in Fig.~\ref{fig:correlations}(a).
The panels (b)-(d) display $\sigma^{\prime}(\omega)$ for \I\ at $T=15$~K,
\Cl\ at $T=15$~K, and \Br\ at $T=20$~K
with separate contributions and the overall fits according to the Drude-Lorenz model.
For all three compounds, two Lorentzians and a couple Fano modes (not shown in the plots) are needed to obtain a satisfactory fit in the mid-infrared spectral range. These contributions correspond to intraband transitions, i.e.\ the transitions within the conduction band split into the lower and upper Hubbard bands ($L_{\rm Hubbard}$); and interband transitions between the dimer bands (intradimer charge transfer, $L_{\rm dimer}$) \cite{PhysRevB.89.205106}. Fano contributions describe the vibrational modes, which are activated due to emv coupling \cite{PhysRevB.76.165113,Dressel2004}. In addition, a Drude peak was added in the case of \Br\ to account for the contribution of the conduction electrons, and one extra Lorentzian term for \I\ (L$_{\rm in\text{-}gap}$), which is ascribed to the realization of localized insulating state with a partially filled gap \cite{PhysRevLett.101.206403}.

\begin{table}
        \caption{Electronic parameter of the Hubbard model extracted from fits of the low-temperature optical conductivity spectra in Fig.~\ref{fig:correlations}. The mid-infrared peak corresponds to the Coulomb repulsion $U$. The electronic bandwidth $W$ is determined by half of the full width of the absorption band. From these experimental values we calculate the correlation strength $U/W$ of $\kappa$-(BEDT-TTF)$_2$Cu[N(CN)$_2$]$X$.}
\label{table:2}
    \begin{center}
    \begin{tabular}{c| c c c}
      $X$ =   &  Cl  & Br &  I \\
    \hline
            $U$ (meV)~~ & 289 & 264 & 294    \\
            $W$ (meV)~~~& 161 & 152 & 132    \\
            $U/W$~~~~ & ~~1.78~~ & ~~1.40~~ & ~~2.20~~    \\
    \end{tabular}
    \end{center}
\end{table}
Most important, from the peak position of $L_{\rm Hubbard}$ we can extract the Coulomb repulsion, while the half-width corresponds to the bandwidth. The values for effective correlation strength $U/W$ obtained from our fits of the optical data $\sigma^{\prime}(\omega)$ \cite{PhysRevB.86.245103} are listed in Tab.~\ref{table:2}.
For \Cl\ and \Br\ the values extracted from the fits of our spectra are higher than those obtained from {\it ab-initio} density functional theory (DFT)
and extended Hückel calculations \cite{PhysRevLett.103.067004}.
We explain this by the sizable renormalization of the bandwidth due to electronic correlations;
in general the experimentally obtained values are  larger than the ones calculated by DFT.
The decrease of the correlation strength $U/W$ when going from \Cl\ to \Br\
corresponds to the common picture of a Mott insulator, on the one hand, and a Fermi liquid
on the other hand that becomes even superconducting at low temperatures \cite{PhysRevB.79.195106,Pustogow2018,Dressel20};
by applying a small amount of pressure enhances the bandwidth sufficiently to cross the insulator-metal transition.
This picture is confirmed by calculations, too.
Along these lines, for \I\ one expects metallic behavior, too, because the size of the anion is bigger. However,
the correlation strength obtained from the optical spectrum is significantly higher. Hence, we have to place
the compound deep into the insulating side of the phase diagram, even beyond the Cl-salt.
Our findings are in accordance with larger dc resistivity and previous studies \cite{PhysRevB.100.195115,PhysRevB.62.15561}.

\subsection{Disorder}
The electronic properties of these molecular conductors are not solely determined by correlation strength;
also disorder has an important influence. In a series of papers, Sasaki and collaborators
showed that x-ray irradiation of $\kappa$-Cl and $\kappa$-Br
mainly affects the heavy Cu atoms  introducing disorder in the anion layers \cite{Sasaki2012,cryst2020374, Yoneyama2010}. This can be monitored by infrared studies because
the vibration modes related to the dicyanamide groups coordinated by the Cu
atoms decrease in intensity upon irradiation.
The defects created in the anions layers cause a random potential modulation that also affects BEDT-TTF layers.

In the present case of $\kappa$-(BEDT-TTF)$_2$Cu[N(CN)$_2$]I, the crystals are of highest quality, pristine and not irradiated; hence the source of disorder is distinctively different. Let us consider the interaction between the anion chains and the BEDT-TTF dimers:
It is well know that the terminal ethylene groups are disordered between eclipsed (tilted in the same direction) and staggered conformations (tilted in opposite directions) at room temperature in all three salts, $\kappa$-Cl, Br, and I. Importantly, however, is the different behavior observed upon cooling.
Pouget {\it et al.} pointed out \cite{pouget2018donor} that it strongly depends on the donor$\cdots$donor and donor$\cdots$anion interactions whether the BEDT-TTF molecules adopt eclipsed or staggered conformations.
When looking at the first two salts --~$\kappa$-Cl and $\kappa$-Br~--
both conformations of the ethylene end-groups are present at room temperature,
with a tendency towards the eclipsed conformation ($83\%$ for \Cl\ and $67\pm 2\%$ for \Br) \cite{Hiramatsu2015,PhysRevB.75.104512}. With lowering the temperature, the eclipsed conformation strongly dominates: for $\kappa$-Cl, the end-groups are completely eclipsed below 150~K;
while for $\kappa$-Br, 97\%\ of the BEDT-TTF molecules possess ethylene groups in the eclipsed conformation.
In other words, the CH$_2$ groups are basically ordered in eclipsed configuration at low temperatures
and only $3\%$ of the ethylene groups are tilted in opposite directions (staggered conformation).
The stabilization of the eclipsed conformation in \Cl\ and \Br\ results from inter-dimer interactions,
which dominate over intra-dimer and donor$\cdots$anion interactions.
For the former salt the CH$\cdots$HC contacts are more strained in the staggered conformation
than in case of \Br, and they become even shorter upon cooling.
Therefore no staggered conformation remains in $\kappa$-Cl at low temperatures
due to destabilizing repulsive interactions.
For $\kappa$-Br these interactions are weaker, and some of the end-groups are still staggered at the lowest temperature \cite{pouget2018donor}.

In contrast to \Cl\ and  \Br, the behavior is rather different for $\kappa$-I because
the terminal ethylene groups stay disordered even at low temperature (70\% in eclipsed and 30\% in staggered conformations). This behavior is mainly governed by  donor$\cdots$donor interactions:
while the eclipsed arrangement leads to strongly strained  CH$\cdots$S contacts, the staggered one
results in strongly strained CH$\cdots$HC contacts. In order to reduce the steric strain in the lattice, the ethylene groups remain disordered even at low temperatures \cite{GEISER1991475}. This disorder causes a random potential that affects the conducting-BEDT-TTF layers in $\kappa$-I. It also prevents superconductivity in \I\ at ambient pressure, very similar to the findings in $\beta$-(BEDT-TTF)$_2$I$_3$ \cite{GEISER1991475,Whangbo1987,PhysRevB.37.5113,PhysRevB.33.7823}. It is interesting to note that the donor$\cdots$anion contacts are shorter in $\kappa$-I compare to the Cl  and Br-analogues \cite{GEISER1991475}.

Although the origin of disorder is different in pristine \I\ with respect to irradiated \Cl\ and \Br,
in all cases randomness leads to Anderson localization and incoherent transport.
To quantify the effect of disorder, let us compare the optical spectra. They provide an integral and energy-resolved property that is very sensitive to the amount of disorder introduced by x-ray irradiation and it can be quantified by the total time of irradiation \cite{cryst2020374,PhysRevLett.101.206403,Sasaki2012}.
It was shown for the insulator \Cl\ that upon increasing the irradiation time there is a significant change of optical conductivity:  the spectral weight shifts from the mid-infrared to the far-infrared.
As a result, the well-developed Mott-Hub\-bard gap gradually closes and fills up, indicating  the localized insulating state \cite{PhysRevLett.101.206403}. When we now compare the low-frequency optical conductivity ($\nu < 1000$~\cm) for \I\ with successively irradiated \Cl\ salts, we can quantify the inherent amount of disorder.

\begin{figure}[t]
\centering
\includegraphics[width=1\columnwidth,clip]{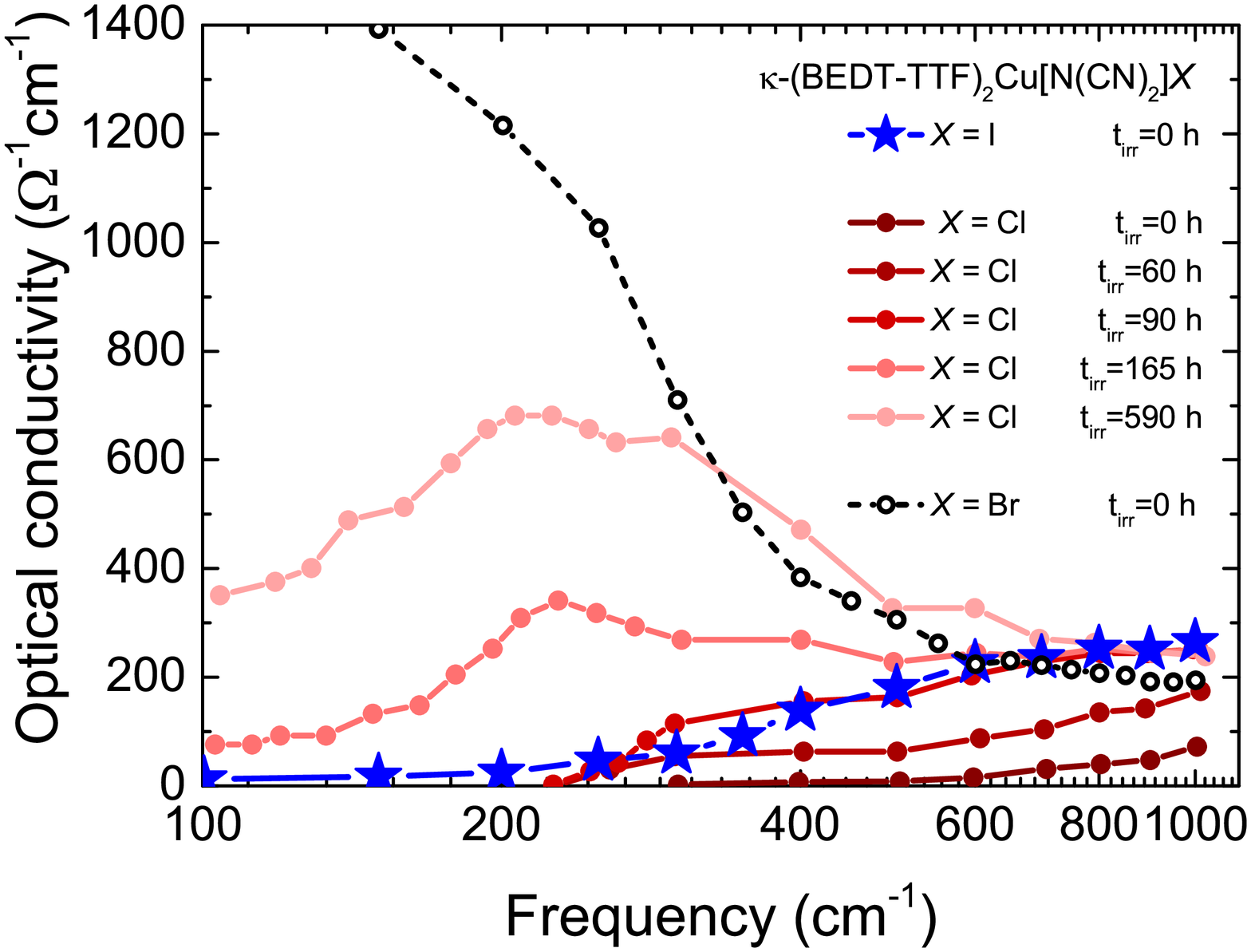}
\caption{Values of optical conductivity of \X\ (X = Cl, I, Br) at different frequencies with different irradiation time ($t_{\rm irr}$) at the lowest temperature ($X=$ Cl data are taken from Ref.~\onlinecite{PhysRevLett.101.206403}). }
\label{fig:disorder}
\end{figure}

To that end, the low-temperature optical conductivity of \I\ is plotted in Fig.~\ref{fig:disorder}
in comparison with the spectra of \Cl\ after being irradiated for different amounts of time as indicated.
With increasing disorder the gap in \Cl\ closes gradually; at around 90 hours of irradiation, the optical behavior
almost coincides with the spectra of \I. We conclude, that in \I\ crystals --~even without irradiation~--  inherent disorder is present, strongly affecting the electronic properties. We trace this effect back to disordered ethylene end-groups of the BEDT-TTF donor molecules.
At elevated temperature, for all three salts, \Cl, \Br\ and \I, dynamical disorder in the terminal ethylene groups dominates with a mixture of the two possible conformations --~staggered and eclipsed. Upon cooling the motion of the -CH$_2$ groups freezes out, and they become ordered in the first two compounds with preferred eclipsed conformation. In the I-salt, however, the ethylene groups remain disordered with both eclipsed and staggered conformations statistically distributed \cite{GEISER1991475, PhysRevB.100.195115}. This random potential causes a partial localization of charge seen in the optical spectra.

Our final results are in line with recent investigations of $\kappa$-(BEDT-TTF)$_2$Cu[N(CN)$_2$]I
by NMR spectroscopy where an
abrupt line broadening below $T=40$~K indicated an electronic inhomogeneity accompanied by antiferromagnetic fluctuations \cite{PhysRevB.100.195115}.
Angluar dependent studies by electron spin resonance spectroscopy also reveal the intrinsic disorder
in the spin behavior of \I\ \cite{PhysRevB.102.214430}.
The transient polarization anisotropy observed by pump probe spectroscopy polarized along the $c$-axis might also be related to the disorder in the terminal ethylene groups affecting the electronic properties \cite{Tsuchiya2020}.

\subsection{Phase diagram}

\begin{figure}[h]
\centering
\includegraphics[width=1\columnwidth,clip]{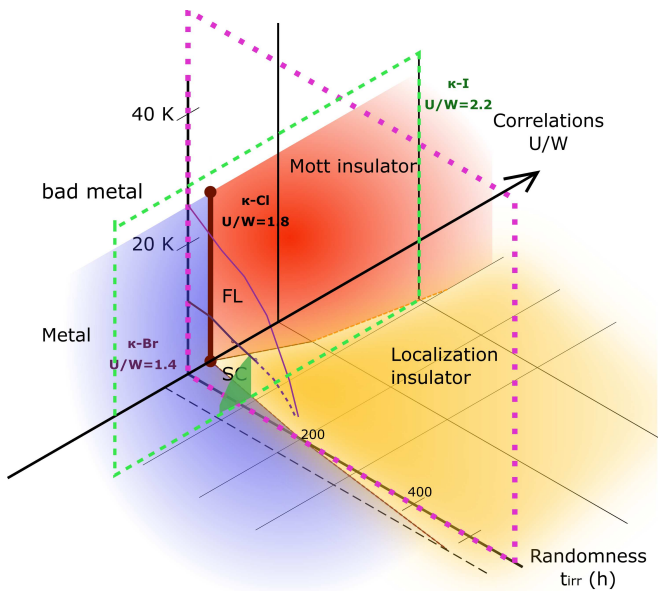}
\caption{Schematic phase diagram of $\kappa$-(BEDT-TTF)$_2$\-Cu[N(CN)$_2$]$X$ ($X$ = Br, Cl, I) for temperature, effective correlation strength ($U/W$), and randomness ($t_{\rm irr}$). The thick brown line indicates the border between metallic and insulating states 
terminated by the critical end point. The plane indicated by the dashed pink line 
shows how the electronic state of $\kappa$-(BEDT-TTF)$_2$Cu[N(CN)$_2$]Br is modified when disorder is introduced;
the superconducting (SC) and Fermi-liquid (FL) states are also marked \cite{cryst2020374}. 
The position of $\kappa$-(BEDT-TTF)$_2$Cu[N(CN)$_2$]Cl is shown by the vertical black line at $t_{\rm irr}$ = 0~h. 
The plane surrounded by the dashed green line corresponds to the change of the  $\kappa$-(BEDT-TTF)$_2$Cu[N(CN)$_2$]I ground state 
by applying hydrostatic pressure, the superconducting phase is represented by the green area \cite{PhysRevB.65.064516}. 
The solid black line here corresponds to $\kappa$-(BEDT-TTF)$_2$Cu[N(CN)$_2$]I at ambient pressure.}
\label{fig:diagram}
\end{figure}

For summarizing our findings, we propose a schematic phase diagram for \X\ ($X$ = Cl, Br, I) salts in Fig. \ref{fig:diagram} that presents 
the temperature-dependent phases as a function of correlation strength ($U/W$) and randomness ($t_{irr}$). 
The data from our optical studies are complemented with previous reports \cite{cryst2020374}, in order to place $\kappa$-I 
together with the other members of the family; this way we could determine the exact positions of $\kappa$-Cl, Br, and I on the $U/W$ axis.

In the absence of artificial disorder, $\kappa$-Br and Cl are placed close to the Mott transition: on the metallic (superconducting) and Mott insulating side, respectively. Introducing disorder in $\kappa$-Br (the plane indicated by the dashed pink line), firstly the superconducting transition temperature decrease. At some critical value of disorder, the system enters the localized insulating state.
The Mott insulator $\kappa$-Cl, on the other hand, is transformed to a localized insulator only after 60~h of irradiation, as
evidenced by the appearance of an in-gap absorption peak in the optical spectra \cite{PhysRevLett.101.206403}.
In the case of \I, the presence of inherent disorder at low temperature places the compound in the Anderson-type localization insulator state even without externally introduced disorder. Applying hydrostatic pressure (the plane indicated by the green dashed line in Fig.~\ref{fig:diagram}) tunes $\kappa$-I  through the insulator-to-metal transition, with the superconducting state indicated by the green area \cite{PhysRevB.65.064516}.

Just to make it clear, this sketch is gross simplified with only borders between metallic (superconducting), Mott insulating, and localized insulating phases shown. As the Mott transition is the first-order transition, a phase-coexistence region is expected at the border of metal and insulator in the absence of disorder \cite{PhysRevLett.91.016401}. For a high degree of randomness a Griffiths-like phase was suggested recently \cite{griffiths2020,PhysRevLett.102.206403,PhysRevLett.124.046404}. For more detailed exploration of these states, optical and dielectric investigations under the pressure are highly desirable.

\section{Conclusions}
\label{sec:conclusions}

Our comprehensive investigations of the charge transport, dielectric response and infrared behavior of $\kappa$-(BEDT\--TTF)$_2$Cu[N(CN)$_2$]I yield valuable information on its electronic properties that allows us to locate \I\ in a global phase diagram with respect to the other members of the  \X\ family.
When going from \Br\ to \Cl, and finally to \I, the electronic correlations strength increases monotonously, and reaches $U/W=2.2$ for the title compound;
in other words, \I\ is situated deeper in the insulating regime than previously expected. 
This contrasts suggestions based on the atomic radius.
However, \I\ does not represent a clear-cut Mott insulator; even below $T\approx 25$~K there remains a finite density of states near the Fermi level. By comparing the optical spectra we identify \I\ as a Coulomb localized insulator, similar to \Cl\ when severely disordered by x-ray irradiation for a period of $t_{\rm irr}=90$~h.
The  \X\ compounds appear now in a new light, as our findings indicate that not only electronic interactions determine the physical behavior but also the role of disorder is crucial for the understanding of these compounds.
These conclusions are more general and might hold for most correlated electron systems.

\begin{acknowledgements}
We thank Gabriele Untereiner for the indispensable technical assistance. The work was supported by the Deut\-sche Forschungsgemeinschaft (DFG) via DR228/39-3 and DR228/52-1.
\end{acknowledgements}

\appendix
\section{Vibrational modes}
\label{Appendix:Vibrations}
\begin{figure}
\centering
\includegraphics[width=1\columnwidth,clip]{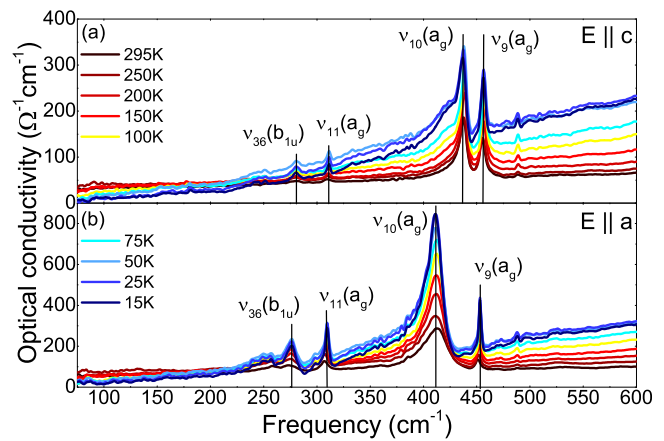}
\caption{In-plane optical conductivity of \I\ in far-infrared spectral range measured for different temperatures 
with the polarization (a) $E \parallel c$ and (b) $E \parallel a$.}
\label{fig:fir}
\end{figure}

Fig. \ref{fig:fir} shows the optical conductivity of \I\ for both in-plane axes recorded in the far-infrared range at different temperatures. The strong vibrational modes at 280, 310, 440 and 460~\cm\ for E$ || $c, and at 275, 309, 409 and 450 \cm\ for E$ || $a are assigned to the $\nu_{36}$(b$_{1u}$), $\nu_{11}$(a$_{g}$), $\nu_{10}$(a$_{g}$) and $\nu_{9}$(a$_{g}$) intra-molecular vibrations of BEDT-TTF \cite{ELDRIDGE1991583,ELDRIDGE1995947,Eldridge1996,Dressel2004}.

\begin{figure}
\centering
\includegraphics[width=1\columnwidth,clip]{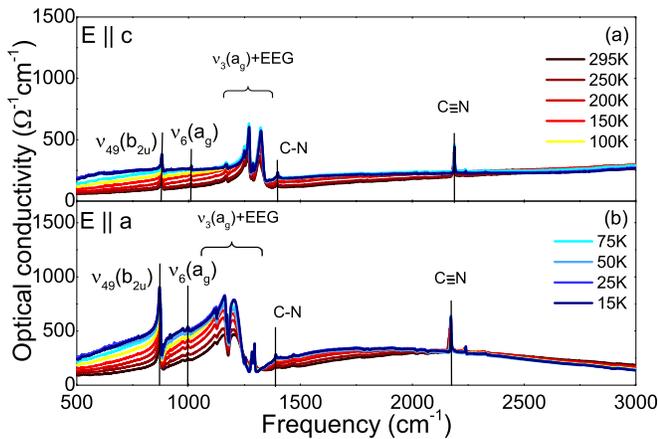}
\caption{Mit-infrared spectra of \I\ along  (a) the  $E \parallel c$ and (b) the $E \parallel a$ polarization for different temperatures.}
\label{fig:mir}
\end{figure}

In the mid-infrared spectral range, for both polarizations the most dominant features
are the totally-symmetric $\nu_3$(a$_g$) vibrations of the C=C double bonds activated via emv coupling (Fig.~\ref{fig:mir}). It appears as a broad resonance between 1250 and 1350 \cm\ for E$ || $c, and between 1100 and 1250 \cm\ for $E \parallel a$. Vibrations  of the ethylene endgroups result in four $\nu_5$(a$_g$) peaks observed at the lower edge of the main resonance for $E \parallel a$,
and antiresonant dips in the $E \parallel c$ spectra \cite{ELDRIDGE1991583,Eldridge1996}. Here the $\nu_3$ mode is shifted to lower frequency compare to $\kappa$-(BEDT-TTF)$_2$Cu[N(CN)$_2$]Br in agreement with higher correlation strength in the former one \cite{sasaki2009}.
In addition, the other strong modes in Fig. \ref{fig:mir} were assigned to $\nu_{49}$(b$_{2u}$), $\nu_6$(a$_g$), and vibrations
related to the dicyanamide group of the anion molecule \cite{Sasaki2012,cryst2020374,jurgens1998}.

\end{document}